\documentclass[12pt]{article}
\usepackage{epsfig}
\usepackage{amsfonts}
\usepackage{latexsym}
\usepackage{amsmath}
\usepackage{comment}
\usepackage[square,comma,sort&compress]{natbib}

\def\half{{1\over 2}}
\numberwithin{equation}{section}

\def\e{{\epsilon}}

 \def\p{\partial}

 \def\ls{\Lambda^2}

\newcommand{\bea}{\begin{eqnarray}}
\newcommand{\eea}{\end{eqnarray}}
\newcommand{\be}{\begin{equation}}
\newcommand{\ee}{\end{equation}}
\newcommand{\non}{\nonumber}
  \makeatletter
  \let\over=\@@over \let\overwithdelims=\@@overwithdelims
  \let\atop=\@@atop \let\atopwithdelims=\@@atopwithdelims
  \let\above=\@@above \let\abovewithdelims=\@@abovewithdelims
  \makeatother
\begin{document}

\begin{titlepage}
\today
\vskip 3 cm
\begin{center}

{}

 {\Large \bf The
Kerr/CFT Correspondence}

\vskip 1.5cm {Monica Guica$^{\dag}$, Thomas Hartman$^\S$, Wei
Song$^{\S\ddag}$, and Andrew Strominger$^\S$}
 \vskip 0.9 cm

{\em \small $^\dag$LPTHE,
Universit\'{e} Pierre et Marie Curie-Paris 6; CNRS\\
 Bo\^{i}te 126, 4 Pl. Jussieu, 75252 Paris Cedex 05, France}
\vskip 0.5 cm
{\em \small $^\S$Center for the Fundamental Laws of Nature\\
Jefferson Physical Laboratory, Harvard University, Cambridge, MA 02138, USA}\\

\end{center}

\vskip 0.6cm

\begin{abstract}
%\vskip 0.5 cm
Quantum gravity in the region very near the horizon of an extreme Kerr black hole (whose angular momentum and mass are related by $J=GM^2$) is considered.  It is shown that consistent boundary conditions exist, for which the asymptotic symmetry generators form one copy of the Virasoro algebra with central charge $c_L={12J \over \hbar}$. This implies that the near-horizon quantum states can be identified with those of  (a chiral half of) a two-dimensional conformal field theory (CFT).
Moreover,
in the extreme limit, the Frolov-Thorne vacuum state reduces to a thermal density matrix
with dimensionless temperature $T_L={1 \over 2 \pi}$ and conjugate energy given by the zero mode generator, $L_0$, of the Virasoro algebra.
Assuming unitarity, the Cardy formula then gives a microscopic entropy
$S_{\rm micro}={2\pi J \over \hbar}$ for the CFT, which reproduces the macroscopic Bekenstein-Hawking entropy
$S_{\rm macro}={{\rm Area }\over 4\hbar G}$. The results apply to any consistent unitary quantum theory of gravity with a Kerr solution. We accordingly conjecture that extreme Kerr black holes are holographically dual to a chiral two-dimensional conformal field theory with central charge $c_L={12J \over \hbar}$,
 and in particular that the near-extreme black hole GRS 1915+105 is approximately dual to a CFT with $c_L \sim 2 \times 10^{79}$.
\noindent

\end{abstract}

\vspace{3.0cm}

\ddag {\it \small On leave from the Institute of Theoretical Physics, Academia
Sinica, Beijing 100080,  China}
\end{titlepage}
\pagestyle{plain}
\setcounter{page}{1}
\newcounter{bean}
\baselineskip18pt

%%%%%%%%%%%%%%%%%%%%%%%%%%%%%%%%%%%%%%%%%%%%%%%%%%%%%%%%%%%%%%%%%%%%%

\tableofcontents
\section{Introduction}
One of the deepest discoveries in modern theoretical physics  is that of holographic dualities, which relate a quantum theory of gravity  to a quantum field theory without gravity in fewer dimensions. These dualities become especially powerful when combined with string theory \cite{Maldacena:1998re}. It is an occasional misconception, however,  that the existence of holographic dualities is contingent on the validity of string theory. This is not the case.  For example, the demonstration \cite{Brown:1986nw} that any consistent theory of quantum gravity on three-dimensional anti-de Sitter space (AdS$_3$) is holographically dual to a two-dimensional conformal field theory (CFT)  did not invoke string theory.  When holographic duality was used to find the microscopic origin of the Bekenstein-Hawking entropy for a class of black holes, the construction at first appeared to depend heavily on details of string theory \cite{ascv}. However, it was later understood \cite{Strominger:1997eq} to apply to essentially any consistent, unitary quantum theory of gravity containing the black holes as classical solutions. In the last few years we are beginning to see interesting applications of holographic duality outside of string theory in nuclear \cite{dragforce,Liu:2006he,Son:2007vk}, condensed matter \cite{Herzog:2007ij,Hartnoll:2007ih,Hartnoll:2008vx} and atomic \cite{balamcgreevy,son} physics.

Oddly, the rich ideas surrounding holographic dualities so far have not been successfully applied to the enigmatic objects which largely inspired their original discovery -- the Schwarzschild or Kerr black holes we actually observe in the sky.\footnote{The successes so far have mainly concerned black holes with large amounts of charge and/or in dimensions other than four.} In this paper we attempt to fill this gap by arguing,
in the spirit of  \cite{Brown:1986nw,Strominger:1997eq}, that extreme Kerr black holes are holographically dual to a chiral CFT in two dimensions.  An extreme Kerr black hole is one for which the angular momentum $J$ saturates the bound $J\le GM^2$. More angular momentum with the same mass $M$ leads to a violation of cosmic censorship. Nearly extreme black holes have been seen in the sky. For example GRS 1915+105, with mass $M\sim 14 M_{sun}$, has $J/GM^2 > 0.98$ \cite{narayan}, and  corrections to the dual CFT representation of GRS 1915+105 should be correspondingly suppressed. In addition, at extremality the ISCO (the innermost stable circular orbit on the accretion disc) coincides with the event horizon, so near extremality the ISCO is within the near-horizon region. Therefore the observed emissions from the ISCO should be well-described by the dual CFT.\footnote{In \cite{jmasa,jmasb} greybody scattering factors for various black holes were computed using the dual CFT picture, and found to agree with those computed by conventional methods. Computations of this type may also be possible for Kerr, and generalized to the context of accretion discs.}  It is our hope that the  rich experimental \cite{narayan,Remillard:2006fc} and theoretical \cite{kth}  literature on Kerr black holes can be illuminated by the dual CFT description.

Our argument that Kerr is dual to a CFT parallels the general one given by Brown and Henneaux \cite{Brown:1986nw} for AdS$_3$, except that we replace AdS$_3$ with the NHEK (near-horizon extreme Kerr) geometry found by Bardeen and Horowitz \cite{bh} via a near-horizon limiting procedure.\footnote{In this procedure the asymptotically flat region, whose excitations we do not regard as part of the black hole itself, is excised and one is left only with the portion of the spacetime neighboring the black hole horizon.} Despite having different dimensions, the spaces bear some resemblance: a slice of NHEK at a particular fixed polar angle is a discrete quotient of AdS$_3$. We first carefully specify boundary conditions at the asymptotic infinity of NHEK (which is where, before taking the near-horizon limit, it is joined to Minkowski space in the full Kerr solution) and demonstrate their consistency.  We then show that, given these boundary conditions, the so-called asymptotic symmetry group (ASG) is one copy of the conformal group and furthermore has a central charge $c_L={12J\over \hbar}$. Hence extreme Kerr, with the given boundary conditions,  is dual to a chiral CFT.\footnote{We do not have an argument for modular invariance and are not distinguishing here between the chiral sector of a nonchiral CFT and a CFT with only a chiral sector. It is interesting to note however that a necessary condition for the partition function of the latter to be modular invariant up to a sign, accounting for the presence of fermions,  given   $c=12J/\hbar$ is precisely that $J/\hbar $ is half-integral.}

While this very general analysis gives the central charge of the dual CFT, it tells us little else about the detailed structure of the CFT. For that to be determined we would need an ultraviolet completion (for example string theory) of quantum gravity on the Kerr background. However the information about the central charge, together with the assumption of unitarity,  turns out to be exactly enough to compute the extreme Kerr entropy by counting quantum microstates, as in
\cite{Strominger:1997eq}. An analysis of the extreme limit of the Frolov-Thorne vacuum, which generalizes the Hartle-Hawking vacuum for Schwarzchild to Kerr, shows that the CFT must be at temperature $T_L={1 \over 2\pi}$. We then apply the thermodynamic Cardy
formula relating the microscopic entropy of a unitary CFT to its temperature and central charge. The resulting entropy agrees exactly with the macroscopic Bekenstein-Hawking area-entropy law, providing corroboration for our proposal that extreme Kerr is dual to a two-dimensional chiral CFT.

The fact that we encounter only a chiral half of a CFT ultimately derives from the fact that at
extremality the rotational velocity of the Kerr horizon becomes the speed of light. Hence both
edges of the forward light cone coincide as the horizon is approached and force
all physical excitations(such as the edge of the accretion disc), which must lie between the edges of light cone,  to spin around chirally with the black hole. Away from extremality
this is no longer the case and we may  expect to encounter
a non-chiral CFT. This very interesting but difficult problem will not be considered herein.

We wish to stress that, while mere consistency imposes very strong constraints, we have not analyzed all possibilities  and have not shown that our near-horizon boundary conditions are the unique consistent choice for studying extreme Kerr. While we did not find any other consistent and nontrivial choices, our search was not exhaustive, and there may well be others with different consequences. Ultimately, the appropriate boundary conditions should be determined from the physical question.
We do suspect that weaker or different boundary conditions will be needed for the just-mentioned problem of
near-extremal excitations. These issues remain for future work.

Section 2 reviews the Kerr geometry and section 3 its near-horizon limit. In section 4 we review the notion of an ASG.  Our boundary conditions are specified in section 5, and the
generators $L_n$ of the corresponding ASG are shown to form a Virasoro algebra  in section 6.  The central charge is computed in section 7. In section 8 we take the limit of the Frolov-Thorne vacuum for Kerr, and show that it yields a thermal state with temperature ${1 \over 2\pi}$. In the concluding section we microscopically compute the entropy for extreme Kerr from  the Cardy formula and find that it reproduces the macroscopic Bekenstein-Hawking area law. Some technical points are relegated to two appendices.

Previous work on a dual description of Kerr, some in the context of string theory, includes \cite{jmasb,Cvetic:1999ja, Horowitz:2007xq,Dabholkar:2006tb}.

\section{Kerr review}

The Kerr metric \cite{krr, Visser:2007fj} is the general rotating black hole solution of the four-dimensional vacuum Einstein equations. In Boyer-Lindquist coordinates it is

\be
ds^2=-{\Delta \over \rho^2}\left(d\hat t-a \sin^2\theta d\hat\phi\right)^2+{\sin^2 \theta \over \rho^2}
\left((\hat r^2+a^2)d\hat \phi-a d\hat t\right)^2+
{\rho^2 \over\Delta}d\hat r^2+\rho^2 d\theta^2
\ee

\be
\Delta\equiv\hat r^2-2Mr+a^2\:,\;\;\;\;\;
\rho^2\equiv\hat r^2 +a^2\cos^2 \theta,\ee

\be a\equiv{G J \over M}\;, \;\;\;\;\; M\equiv GM_{ADM} \ee It is
labeled by two parameters: the angular momentum $J$ and the
geometric mass $M$. In order to simplify the formulae, but at the
risk of some confusion, in the above and hereafter we have rescaled
$M$ by a factor of $G$ relative to the abstract and introduction.
The solution has naked singularities unless $J$ lies in the
parameter range \be {-M^2\over G}\le  J \le {M^2\over G}.\ee Of
course, quantum mechanically $J$ is quantized \be J=\hbar j  \ee for
some half integer $j$. There is an event horizon at \be
r_+=M+\sqrt{M^2-a^2}. \ee The Hawking temperature, surface gravity
and angular velocity of the horizon are \be\label{th} T_H={\hbar
\kappa \over 2\pi}={\hbar(r_+-M) \over 4\pi Mr_+}, \ee \be
\Omega_H={a \over 2Mr_+}. \ee These are related by the first law to
the Bekenstein-Hawking entropy
\cite{Bekenstein:1973ur,Hawking:1975sw} \be S_{BH}={\mbox{Area}
\over 4\hbar G}={2 \pi Mr_+ \over \hbar G}. \ee

We are primarily interested in the so-called extreme Kerr, which carries the maximum allowed angular momentum
\be  J={M^2\over G}. \ee
Extreme Kerr has zero Hawking temperature but a nonzero
entropy
\be\label{sbh}
S_{BH}={2\pi J \over \hbar}
\ee
Our goal is to  to explain this number as the logarithm of the number of quantum microstates of Kerr.

\section{The NHEK geometry}

We wish to study the region very near the extreme Kerr horizon at $\hat r=M$. In order to do so, following Bardeen and Horowitz \cite{bh} we define, new (dimensionless) coordinates

\be
t=\frac{\lambda \hat{t}}{ 2M}\;, \;\;\;\;\;
y= \frac{ \lambda M}{\hat{r}-M} \;, \;\;\;\;\;
\phi=\hat{ \phi}- {\hat{t} \over 2M}
\ee

\vskip 0.4 cm
\noindent and take
$\lambda \to 0$ keeping $(t,y,\phi,\theta)$ fixed. The result is the
near-horizon extreme Kerr or ``NHEK" geometry in Poincar\'e-type
coordinates

 \be\label{nhk} {ds^2 }= 2 G J\Omega^2 \left(
{-dt^2+dy^2\over y^2}
 + d\theta^2 +\ls  (d\phi + {dt\over y})^2\right)
\ee
where

\be
\Omega^2 \equiv {1+\cos^2\theta\over 2}\;, \;\;\;\;\; \Lambda \equiv {2 \sin \theta\over 1+\cos^2\theta},
\ee
\vskip 0.4 cm
\noindent
$\phi \sim \phi + 2 \pi$ and $0\le \theta \le \pi$. The NHEK geometry is not asymptotically flat. Note that the angular momentum affects only the overall scale of the geometry.

The coordinates (\ref{nhk}) cover only part of the NHEK geometry. Global coordinates
$(r,\tau,\varphi)$ are given by (for a discussion of global properties see \cite{bh})

\bea
y&=&\left(\cos \tau\sqrt{1+r^2}+r\right)^{-1},\\
t&=&y \sin \tau\sqrt{1+r^2},\\
\phi&=&\varphi+\ln\left({\cos\tau+r\sin\tau \over
1+\sin\tau\sqrt{1+r^2}}\right). \eea The metric (\ref{nhk}) is then
 \be\label{nhkg}
{d\bar s^2 }= 2 G J\Omega^2 \left( -(1+r^2){d\tau^2} + {dr^2\over 1+r^2} + d\theta^2 + {  \ls}(d\varphi + {rd\tau})^2\right).
\ee

The NHEK geometry has an enhanced $SL(2,\mathbb{R})\times U(1)$ isometry group \cite{bh}. The rotational $U(1)$ isometry is generated by the Killing vector
\be
{\zeta_0}=-\p_\varphi.
\ee
Time translations become part of an enhanced $SL(2,\mathbb{R})$ isometry group generated by the Killing vectors
\begin{eqnarray}
\tilde{J}_1 &=&  {2\sin{\tau} {r\over\sqrt{1+r^2}} \partial_{\tau} - 2\cos{\tau}\sqrt{1+r^2} \partial_{r}+\frac{2\sin{\tau}}{\sqrt{1+r^2}} \partial_{\varphi} }  \\
\tilde{J}_2 &=&  {-2\cos{\tau} {r\over\sqrt{1+r^2}} \partial_{\tau}-2\sin{\tau} \sqrt{1+r^2} \partial_{r} - \frac{2\cos{\tau}}{\sqrt{1+r^2}} \partial_{\varphi}  }  \\
\tilde{J}_0 &=&  2\partial_{\tau}
\end{eqnarray}
Note that all of these isometries act within a three-dimensional
slice of fixed polar angle $\theta$. The geometry of these slices is
a quotient of warped AdS$_3$ (the AdS$_3$ analog of the squashed
S$^3$), with the quotient arising from the $\varphi$ identification
\cite{Detournay:2005fz,Bengtsson:2005zj}. Such quotients are
(warped) black holes, much as AdS$_3$ quotients are BTZ black holes
\cite{stromrecent}. The $\tau,r$ plane describes AdS$_2$, while the
$\varphi$ circle is an $S^1$ bundle over the AdS$_2$. At the special
value of $\theta$ where $\Omega^2=\sin \theta$, the slice is locally
an ordinary AdS$_3$ and acquires a local $SL(2,\mathbb{R})_R\times
SL(2,\mathbb{R})_L$ isometry.  At all other values of $\theta$, the
$SL(2,\mathbb{R})_L$ is broken to U$(1)$. Near the equator we have a
``stretched" AdS$_3$ quotient (as the $S^1$ fiber is stretched),
while near the poles we have a ``squashed" AdS$_3$ quotient.
Properties of these three-dimensional spacetimes in a context
relevant to the present one were recently described in
\cite{stromrecent}.

\section{The Asymptotic Symmetry Group}

 We now turn to the study of excitations around  near-horizon extreme Kerr. This requires imposing boundary conditions at the $S^2\times \mathbb{R}$ boundary $y=0$.  Since we lost the asymptotically flat region in taking the near-horizon limit, this boundary is not flat and it is not a priori obvious what boundary conditions we should use. Indeed, different boundary conditions may be relevant in different physical contexts.
 For every consistent set of boundary conditions there is an associated asymptotic symmetry
group (ASG). This is defined as
the set of allowed symmetry transformations modulo
the set of trivial symmetry transformations
\medskip

\begin{equation}{\rm ASG}= \frac{\rm Allowed~~Symmetry~~Transformations}
{\rm Trivial~~Symmetry~~Transformations}.
\end{equation}
\vskip 0.4 cm
\noindent
Here `allowed' means
that the transformation is consistent with the specified boundary conditions, while `trivial'
means that the generator of the transformation vanishes after we
have implemented the constraints and reduced it to a boundary integral.

Consistency requires that the generators of the ASG be well defined and  not diverge at the boundary. If the boundary conditions are too strong, all interesting excitations are ruled out. If they are too weak, the generators of the ASG are ill-defined. In general, there is a narrow window of consistent boundary conditions. For example, in asymptotically flat space, one usually requires that excitations of the metric fall off like ${1 \over r}$ or faster at infinity. The ASG is then simply the Poincar\'e group. One might try to demand
that the metric fall off spatially as ${1 \over r^2}$. This would
allow only zero energy configurations and hence the theory would
be trivial. On the other hand, one might try to demand that it fall
off as ${1 \over \sqrt{r}}$. Then the energy and other symmetry
generators would be in general divergent, and it is unlikely any sense could
be made of the theory. So the general idea is to make the falloff weak enough to
include the physics of interest, while still maintaining finiteness of the generators.

\section{Boundary Conditions}
We choose the boundary conditions
\be\label{strictbc}
\left(
  \begin{array}{ccccc}
 h_{\tau\tau}= \mathcal{O}({r^2}) & h_{\tau\varphi}= \mathcal{O}({1}) & h_{\tau\theta}= \mathcal{O}({1\over r}) &h_{\tau r}= \mathcal{O}({1\over r^2})  \\
 h_{\varphi \tau}=h_{\tau\varphi} & h_{\varphi\varphi}= O(1) &h_{\varphi\theta}= \mathcal{O}({1\over r})  &h_{\varphi r}= \mathcal{O}({1\over r})  \\
   h_{\theta \tau}=h_{\tau\theta} & h_{\theta\varphi}=h_{\varphi\theta} & h_{\theta\theta}= \mathcal{O}({1\over r}) &h_{\theta r}= \mathcal{O}({1\over r^2}) \\
   h_{r\tau}=h_{\tau r} & h_{r\varphi}=h_{\varphi r} & h_{r\theta}=h_{\theta r} & h_{rr}= \mathcal{O}({1\over r^3}) \\
  \end{array}
\right)\ , \ee
where $h_{\mu\nu}$ is the deviation of the full metric from the background NHEK metric $\bar g$ in (\ref{nhkg}).
We note that the allowed deviations $h_{\tau\tau}$ and $h_{\varphi\varphi}$\footnote{The asymptotic constraints force a linear combination of these, the trace of $h_{\mu\nu}$, to vanish at linear order, as described in appendix A.} are of the
same order as the leading terms in (\ref{nhkg}). In this regard, these boundary
conditions differ  for example from the usual AdS$_3$ boundary conditions \cite{Brown:1986nw}, where all
deviations are subleading. An analysis with a number of similarities to the present one (with non-subleading  deviations ) for the BMS group at ${\cal I}^+$ can be found
in \cite{bms,barnichbms}.
The most general diffeomorphisms which preserve the
boundary conditions  (\ref{strictbc}) are of the form \be\label{allw} \xi = [-r
\epsilon'(\varphi) + O({1})]\p_r + [C+O({1\over r^3})]\p_\tau
 + [\epsilon(\varphi) + O({1\over r^2})]\p_\varphi + O({1\over r})\p_\theta
\ee where $\epsilon(\varphi)$ is an arbitrary smooth function of the boundary
coordinate $\varphi$, and C is an arbitrary constant. The subleading
terms indicated above can be seen, after computing the generators,
to correspond to trivial diffeomorphisms. Therefore the asymptotic
symmetry group contains one copy of the  conformal group of the
circle generated by\footnote{$\zeta_\epsilon$ is discontinuous at the north and south poles $\theta=(0,\pi)$. This can be regulated by taking for example $\tilde \zeta_\epsilon ={r^2\sin\theta \over 1+r^2\sin \theta} [\epsilon(\varphi)\p_\varphi-r \epsilon'(\varphi)\p_r]$. Expanding in $1 \over r$ we see that  $\tilde \zeta_\epsilon$
and $\zeta_\epsilon$ differ by trivial diffeomorphisms, while $\tilde \zeta_\epsilon$ is smooth for any finite $r$.} \be\label{leftvir} \zeta_\epsilon =
\epsilon(\varphi)\p_\varphi-r \epsilon'(\varphi)\p_r \ . \ee This Virasoro algebra here has only a $U(1)$, not
an $SL(2,\mathbb{R})$, isometry subgroup.\footnote{ This suggests that the CFT
state dual to the Kerr vacuum is not $SL(2,\mathbb{R})$ invariant. } The NHEK
 metric (\ref{nhk}) transforms under (\ref{leftvir}) as

\be\label{ling}
 \delta_\epsilon d\bar s^2=4JG\Omega^2\left(r^2{(1-\ls)\p_\varphi \epsilon } d\tau^2-{r\p_\varphi^2\epsilon \over 1+r^2}d\varphi dr+\ls{\p_\varphi \epsilon }d\varphi^2-{\p_\varphi\epsilon\over (1+r^2)^2}dr^2\right) .
 \ee
\vskip 0.3 cm \noindent Since $\varphi \sim \varphi + 2\pi$ (because
$\phi \sim \phi + 2\pi$ ), it is convenient to define
$\epsilon_n(\varphi) = -e^{-i n \varphi}$ and
$\zeta_n=\zeta(\epsilon_n)$. Under Lie brackets, these symmetry
generators obey the Virasoro algebra \be i[\zeta_m, \zeta_n]_{L.B.}
= (m-n)\zeta_{m+n} \ . \ee Note that $\zeta_0$ generates the $U(1)$
rotational isometry.

The allowed symmetry transformations (\ref{allw}) also include $\tau$
translations generated by $ \p_\tau$.
    The conjugate conserved quantity, which we denoted $E_R$, measures the deviation
${M^2 \over G}-J$ of the black hole from extremality.  Here we wish to study only the extremal black holes, which entails a restriction to the subspace in which $E_R$ vanishes. This restriction should be compatible with (\ref{leftvir}) because $\p_\tau$ commutes with the Virasoro generators.  It can be implemented with an additional boundary condition, given in section \ref{s:generators}  below, which makes the generator of $\tau$ translations trivial.

The reader may wonder how we came up with  the boundary conditions (\ref{strictbc}). We began by assuming (a) the existence of a non-trivial Virasoro whose zero mode is proportional to $\p_\varphi$ in the allowed diffeomorphisms, (b) the boundary conditions can be linearly described in terms of power law falloff of the individual components of the metric fluctuations.  We found only one self-consistent set of boundary conditions with these properties, up to possible further constraints on subleading terms which do not affect the ASG or its central charge. In studies of the G\"{o}del black hole \cite{comperedetournay} and warped AdS$_3$ \cite{Compere:2008cv}, consistent  boundary conditions were imposed in which the $SL(2,\mathbb{R})$ isometry is enhanced to a Virasoro algebra, and the $U(1)$ isometry is enhanced to a current algebra. That is quite different than the situation here (as well as in \cite{mgas}) in which the $SL(2,\mathbb{R})$ becomes trivial and the $U(1)$ is enhanced to a Virasoro and therefore do not meet requirement (a) above. We expect that consistent boundary conditions analogous to those described in \cite{comperedetournay,Compere:2008cv} do exist for Kerr. If so, they are likely relevant to an understanding of the entropy of near-extremal fluctuations (since the $\bar L_0$ of the $SL(2,\mathbb{R})$ measures the deviation from extremality) rather than the ground state entropy of extreme Kerr.

\section{Generators}\label{s:generators}
Now we need to construct the surface integrals which generate
the diffeomorphisms of (\ref{leftvir}) via Dirac brackets, and see if they are finite.  When the deviations $h$ of the metric are not subleading, the charges can have
nonlinear corrections, which must be carefully considered. For this purpose the covariant formalism of Barnich, Brandt and Comp\`{e}re \cite{barnichbrandt,barnichcompere}, based on \cite{abbottdeser,iyerwald,andersontorre,torre,bbheneaux, bbheneaux2} and  further developed in \cite{barnichstokes,comperethesis}, is the most complete and will be adopted in the following.   An example, mathematically quite similar to the present one, are the G\"{o}del black holes analyzed in \cite{comperedetournay}.

The generator of a diffeomorphism $\zeta$ is a conserved charge
$Q_\zeta[g]$.\footnote{We choose the arbitrary additive constants
appearing in \cite{barnichbrandt,barnichcompere} so that
$Q_\zeta[\bar g]=0$  for $\bar g$ the NHEK metric.} Under Dirac
brackets, the charges associated with asymptotic symmetries obey the
same algebra as the symmetries themselves, up to a possible central
term. Infinitesimal charge differences between neighboring
geometries $g_{\mu\nu}$ and $g_{\mu\nu} + h_{\mu\nu}$ are given by
\be\label{infcharge} \delta Q_\zeta[g] = {1\over 8 \pi
G}\int_{\p\Sigma} k_\zeta[h,g] \ee where the integral is over the
boundary of a spatial slice and \bea\label{chargeintegrand}
k_\zeta[h,g]&=& -{1\over 4} \epsilon_{\alpha\beta\mu\nu}\big[
\zeta^\nu D^\mu h - \zeta^\nu D_\sigma h^{\mu\sigma} + \zeta_\sigma D^\nu h^{\mu\sigma} + \half h D^\nu \zeta^\mu \\
& & - h^{\nu\sigma} D_\sigma \zeta^\mu + \half
h^{\sigma\nu}(D^\mu\zeta_\sigma +
D_\sigma\zeta^\mu)\big]dx^\alpha\wedge dx^\beta \,.\notag \eea
Covariant derivatives and raised indices are computed using
$g_{\mu\nu}$.  In asymptotically AdS spacetimes, the formula
(\ref{infcharge}) for the charge is true even for finite $h$, and it
agrees with the charges obtained in the classic Hamiltonian
\cite{reggeteitelboim,Brown:1986nw,barnichads} or quasilocal
\cite{Brown:1992br,krausbala} formalisms. However, in certain cases
such as 5d G\"{o}del spacetimes \cite{barnichgodel,godelbanados},
nonlinear contributions are important near the boundary, and only
infinitesimal $h$ is allowed. In those cases, finite charge
differences are computed by integrating $\delta Q$ over a path in
the configuration space, \be\label{gfz} Q_\zeta[g] - Q_\zeta[\bar g]
= \int_\gamma \delta Q_\zeta[g(\gamma)] \ee where $\gamma$ connects
$\bar g$ to $g$ and $h(\gamma)$ in (\ref{infcharge}) is taken
tangent to the path. Path-independence holds provided certain
integrability conditions are satisfied
\cite{barnichcompere,comperethesis}. We show that these conditions
are obeyed around NHEK in appendix B.

The charges that generate $\p_\tau$  and $\zeta_\epsilon$ are \be
Q_{\p_\tau}= {1\over 8 \pi G}\int_{\p\Sigma} k_{\p_\tau},
\hspace{1cm} Q_{\zeta_\epsilon} = {1\over 8 \pi G}\int_{\p\Sigma}
k_{\zeta_\epsilon}. \ee Choosing $g_{\mu\nu}$ to be the NHEK
background, the integrands simplify to \bea k_{\p_\tau} &=&
-({1\over4\Lambda}r[ (\Lambda^4+\ls -2)h_{\varphi\varphi}
+{\Lambda^4\over r^2}h_{tt}]\\ \non &&-{1\over
4\Lambda}[r^3\Lambda^4h_{rr}+2r^2\Lambda\p_\theta(\Lambda
h_{r\theta})+2\Lambda^2r\p_\tau h_{r\varphi}+2(\ls -1)r^2\p_r
h_{\varphi\varphi}\\
\non&&+2\Lambda^4 h_{\tau\varphi}-\Lambda^2r(\ls -2+2r\p_r)h_{\theta\theta}] )d\theta \wedge d\phi +\cdots\\
k_{\zeta_\epsilon}&=&{1\over4\Lambda}[2\ls \epsilon' r h_{r\varphi}
- \epsilon \ls (\ls {h_{\tau\tau}\over r^2} + (\ls  +
1)h_{\varphi\varphi} + 2 r \p_\varphi h_{r\varphi}) ]d\theta \wedge
d\varphi+\cdots \eea We have assumed the boundary conditions
(\ref{strictbc}) and discarded total $\varphi$ derivatives. The
$+\cdots$ includes terms which vanish for   $r\rightarrow\infty$ or
are not tangent to $\p\Sigma$, and so  do not contribute to the
integral.  From the boundary conditions (\ref{strictbc}) we see
immediately that $k_{\zeta_\epsilon}$, and therefore
$Q_{\zeta_\epsilon}$, are finite around NHEK. For a general
background $g_{\mu\nu}$, a straightforward counting of powers of $r$
term by term in (\ref{chargeintegrand}) reveals that
$Q_{\zeta_\epsilon}$ remains finite.

In addition, we must show that $Q_{\p_\tau}$, which measures the deviation from extremality,  is well-defined. This does not follow immediately from the boundary conditions (\ref{strictbc})\footnote{A similar structure was encountered in \cite{comperedetournay}, who similarly impose a supplementary boundary condition.}.  In fact, as we are studying extreme Kerr,  we want this charge not only to be finite, but to vanish altogether, i.e. to be trivial. We therefore  impose  the supplementary boundary condition
\be\label{rst}
E_R\equiv Q_{\p_\tau}[g]=0.
\ee
This is equivalent to requiring  that the pullback of $k_{\p_\tau}$ to the boundary obeys $k_{\p_\tau}|_{\p\Sigma}=d X|_{\p\Sigma}$
for some one-form $X$ globally defined on $\p\Sigma$.   Under the constraint (\ref{rst}), only perturbations $h$ which preserve  (\ref{rst})  and only background metrics $g$ which can be reached from the NHEK geometry via  a path of such perturbations are considered. This is presumably a complicated nonlinear submanifold of the
geometries allowed by the linear boundary conditions (\ref{strictbc}). It can be shown that the $E_R=0$ submanifold contains in particular finite generalizations of the infinitesmal  $\zeta_\epsilon$ diffeomorphisms acting on  the NHEK geometry.\footnote{Verifying this by explicit computation is a bit tricky  because of subtleties at the north and south pole, and uses the fact that $dk_{\p_\tau}=0$ on shell \cite{barnichbrandt}.  To make the computation well defined, one must use a regulated form of $\zeta_\epsilon$ as e.g. given in footnote 6.  } These carry nonzero  $Q_{\zeta_\epsilon}$ charges. The inclusion of such spaces is expected because the $\zeta_\epsilon$ and $\p_\tau$ commute. We do not know if there are other types of spaces with $E_R=0$. The answer likely depends on the matter content of the theory, about which so far we have assumed only that it does not affect the boundary behavior.

It remains to be seen that, with the supplementary boundary condition  (\ref{rst}), the transformations $\zeta_\epsilon$ are still allowed. Formally this follows from the fact that $\zeta_\epsilon$
and $\p_\tau$ commute, but we must be careful about divergences. It is easy to check directly that the perturbation (\ref{ling}), which results from  the action of $\zeta_\epsilon$ on the NHEK geometry, indeed yields a
$k_{\p_\tau}$ obeying (\ref{rst}). For the more general background consistent with (\ref{rst}), we use the fact that the generators $Q_{\zeta_\epsilon}$ are well defined on the bigger space of geometries obeying only (\ref{strictbc}). Therefore, they properly generate the local action of a $\zeta_\epsilon$ diffeomorphism. This  will preserve the local expression $k_{\p_\tau}|_{\p\Sigma}=d X|_{\p\Sigma}$ of  $k_{\p_\tau}$ as an exact form on $\p\Sigma$ up to a c-number corresponding to a possible central term. The central term is \cite{barnichbrandt}
\be\label{erx}
{1 \over 8\pi G}\int_{\p\Sigma}k_{\zeta_\epsilon}[{\cal L}_\tau \bar g,\bar g]
\ee
where ${\cal L}_\tau$ is the Lie derivative along $\tau$.
%On the subspace of interest, where both $Q_{\p_\tau}$ and $Q_{\zeta_\epsilon}$ are well defined,the central term (\ref{erx}) must also be well-defined.
As there is no possible central term between the generators of Virasoro and $\tau$ translations, this must vanish, in agreement with explicit computation. Therefore we can consistently restrict to extremal configurations by imposing (\ref{rst}).

\section{Central Charge}
The Dirac bracket algebra of the asymptotic symmetry group is
computed by varying the charges
\be \{Q_{\zeta_m},Q_{\zeta_n}\}_{D.B.} =Q_{[\zeta_m,\zeta_n] }+
{1 \over 8\pi G}\int_{\p\Sigma}k_{\zeta_m}[{\cal L}_{\zeta_n} \bar g,\bar g].
\ee
For the NHEK geometry the Lie derivative gives \bea
{\cal L}_{\zeta_n} \bar g_{\tau\tau} &=& {4GJ\Omega^2(1-\ls)}r^2in e^{-in\varphi}\\
{\cal L}_{\zeta_n}\bar  g_{r\varphi} &=& -{2GJ\Omega^2r\over 1+r^2}n^2 e^{-in\varphi}\\
{\cal L}_{\zeta_n}\bar  g_{\varphi\varphi} &=& {4GJ\ls \Omega^2}i n
e^{-i
n\varphi}\\
{\cal L}_{\zeta_n} \bar g_{rr} &=& -{4GJ\Omega^2\over (1+r^2)^2}in
e^{-in\varphi} \eea It follows that \be {1 \over 8\pi
G}\int_{\p\Sigma}k_{\zeta_m}[{\cal L}_{\zeta_n}
\bar g,\bar g]=-i(m^3+2m)\delta_{m+n}{J} \ee Let us now define dimensionless
quantum versions of the $Q$s by
 \be \hbar L_n \equiv
Q_{\zeta_n}+{3J\over2}\delta_{n}, \ee plus the usual rule of Dirac
brackets to commutators as $\{.,.\}_{D.B.}\to-{i \over \hbar}[.,.]$.
The quantum charge algebra is then  \be [L_m, L_n]= (m-n) L_{m+n} +
{J\over\hbar}m(m^2 - 1)\delta_{m+n,0}. \ee From this  we can read
off the central charge for extreme Kerr \be\label{cc} c_L =
{12J\over \hbar } \ . \ee For GRS 1915+105, this gives $c_L=(2\pm
1)\times 10^{79}$, with the uncertainty coming from the uncertainty
in the measured mass.

We note that (\ref{cc}) does not depend on the details of the boundary conditions
(\ref{strictbc}) in that it holds for any boundary conditions as long as the diffeomorphisms (\ref{leftvir}) are allowed.

\section{Temperature}
In this section we derive the relation $T_L={1 \over 2\pi}$ for the generalized temperature of the near-horizon region in units of its inverse radius.

First, we must define the quantum vacuum for extreme Kerr. This problem i
s subtle because Kerr has no everywhere timelike Killing vector, so in fact, globally, there is no quantum state with all the desired properties of a vacuum. There is an extensive literature on this subject for the generic Kerr black hole, references to which can be found in \cite{Duffy:2005mz}. Frolov and Thorne \cite{Frolov:1989jh} define a vacuum by using a Killing vector field which is timelike from the horizon out to the speed of light surface, which is  the surface at which an observer must move at the speed of light in order to corotate with the black hole. The Frolov-Thorne vacuum  has some pathologies outside of this surface \cite{Ottewill:2000qh}, but is well behaved in the near-horizon region \cite{Duffy:2005mz}, where it is an analog of the Hartle-Hawking vacuum for Schwarzschild and is therefore ideal for our purposes.

Construction of the Frolov-Thorne vacuum for generic Kerr begins by expanding  the quantum  fields in eigenmodes of the asymptotic energy $\omega$ and angular momentum $m$. For example for a scalar field $\Phi$ we may write
\be
\Phi= \sum_{\omega, m,l} \phi_{\omega ml} e^{-i\omega \hat t+i m\hat  \phi}f_l(r,\theta).
\ee
After tracing over the region inside the horizon, the vacuum is a diagonal density matrix in the energy-angular momentum eigenbasis with a Boltzmann
weighting factor
\be\label{ggf}
e^{-\hbar \frac{\omega-\Omega_{H} m}{T_H} }.
\ee
This reduces to the Hartle-Hawking vacuum in the non-rotating $\Omega_H=0$ case.

In order to transform this to near-horizon quantities and take the
extremal limit (in which $T_H\to 0$) we note that in the
near-horizon coordinates
\be e^{-i\omega \hat t+i m
\hat \phi}=e^{-{i\over \lambda}(2M\omega -{m}) t+i m\phi}=
e^{-in_Rt+i n_L \phi}, \ee where \be n_L\equiv m,~~~~~~~~n_R\equiv
{1 \over \lambda}(2M\omega -{m}) \ee
are the left and right
charges associated to $\p_\phi$ and $\p_t$ in the near-horizon
region. In terms of these variables the Boltzmann factor (\ref{ggf})
is \be e^{-\hbar{\omega-\Omega_H m\over T_H} }=e^{-{n_L\over T_L}
-{n_R \over T_R}}, \ee where the dimensionless left and right
temperatures are \be T_L={  r_+-M\over 2\pi
(r_+-a)},~~~~~T_R={r_+-M \over2\pi \lambda r_+}. \ee In the
extremal limit $M^2\to GJ$ these reduce to \be T_L={1 \over
2\pi},~~~~~T_R=0. \ee The left-movers are then thermally populated
with the Boltzmann distribution at temperature $1/2\pi$:\footnote{A
fast but less rigorous way to derive this result is to note that at
every fixed polar angle $\theta$, the geometry is a quotient of
warped AdS$_3$. The temperature for such quotients is the length of
the shift determining the quotient divided by $4\pi^2$
\cite{stromrecent,Maldacena:1998bw}. This gives $T_L={1 \over 2
\pi}$ for every  $\theta$.} \be e^{-2\pi n_L}. \ee Note that even
though extreme Kerr has zero Hawking temperature, the quantum fields
outside the horizon are not in a pure state.

\section{Microscopic origin of the Bekenstein-Hawking-Kerr entropy}

In the previous section we saw that the quantum theory in the
Frolov-Thorne vacuum restricted to extreme Kerr has the left-moving
temperature \be\label{tpi} T_L={1 \over 2 \pi}. \ee
 Since the states
of quantum gravity on NHEK, with the boundary conditions
\eqref{strictbc}, are identified under the holographic duality with
those of the left-moving part of the CFT, the CFT dual of the
Frolov-Thorne vacuum must also have temperature \eqref{tpi}. The
central charge of the CFT was shown to be \be\label{ccb} c_L={12 J
\over \hbar}. \ee According to the Cardy formula the entropy for a
unitary CFT at large $T_L$ obeys\footnote{A sufficient but not
necessary condition for validity of the Cardy formula is $T>> c$.
This condition is not obeyed here, as in many black hole
applications \cite{ascv}. In many such cases the formula is
nevertheless valid because of the small gap arising from highly
twisted sectors \cite{Maldacena:1996ds}. For example we might expect
a twisted sector of order $J$, which is effectively described by a
universal $c_L=12$ "long string" CFT at temperature $T_L={J \over
2\pi}$. A small gap is generic for black holes
\cite{Preskill:1991tb} so we hope that the same mechanism is
operative here.} \be S = {\pi^2\over 3}c_L T_L. \ee Using
\eqref{tpi},\eqref{ccb}, we find the microscopic entropy for the dual
to extreme  Kerr \be S_{micro}={2 \pi J \over \hbar}=S_{BH}. \ee
This exactly reproduces the macroscopic Bekenstein-Hawking entropy
(\ref{sbh}) of the extreme Kerr black hole.

\section*{Acknowledgements}

This work was supported in part by DOE grant DE-FG02-91ER40654. We
are grateful to Dionysios Anninos, Geoffrey Comp\`ere, Allison
Farmer, Valeri Frolov, Gary Horowitz and Greg Moore for useful
conversations. W. S. thanks the High Energy Group at Harvard for
their kind hospitality.

\appendix

\section{Asymptotic constraints}
\setcounter{equation}{0}  % reset counter

In this appendix we work out the asymptotic form of the constraint equations, which relate the
leading order fluctuations of the metric.  In a Dirac bracket formalism, the constraints, by construction, commute with everything.
Therefore the generators of the ASG are ambiguous up to the additions of integrals
proportional to the constraints.

The constraint equations are $G^0_\mu = 0$.  Using the boundary conditions (\ref{strictbc}), linearizing in $h_{\mu\nu}$ and expanding to leading order in $1/r$, we can solve the asymptotic constraint equations as follows.

First consider $G^0_\varphi = 0$.  At leading order this is a second
order differential equation for $h_{\varphi\varphi}$ in $\theta$,
and does not involve the other metric components.  The solution
which leads to a metric regular at the poles is \be\label{hpp}
h_{\varphi\varphi} = \Lambda^2\Omega^2 f(\tau, r,\varphi)\ . \ee Now
consider $G^0_0 = 0$. This is a function only of $\theta,
h_{\tau\tau}, h_{\varphi\varphi}$, and their first and second
$\theta$-derivatives.  Plugging in the solution for
$h_{\varphi\varphi}$, all the derivatives drop out and the solution
is \be\label{htt} h_{\tau\tau} = {r^2}(1-\Lambda^2)\Omega^2
f(\tau,r,\varphi)+O(r)\ . \ee

Now consider $G^0_\theta = 0$.  This is proportional to \be
2\Lambda^2 (\Lambda\p_\theta\Omega -
\Omega\p_\theta\Lambda)h_{r\varphi}-\Lambda^3\Omega\p_\theta
h_{r\varphi} - \Omega\p_\theta\Lambda \p_\varphi h_{\varphi\varphi}
\ee Plugging in the solution above for the $\theta$-dependence of
$h_{\varphi\varphi}$, we find \be h_{r\varphi} = -{1\over
r}\left({\Omega^2\over 2}\p_\varphi f(\tau,r,\varphi) +
{16\Omega^2\over\Lambda^2}g(\tau,r,\varphi)\right) + O({1\over r^2})
\ee

Now consider $G^0_r = 0$.  This is a function of $\theta,
h_{r\varphi}, \p_\theta h_{t\varphi}, \p_\theta^2 h_{r\varphi},
\p_\varphi h_{\varphi\varphi}, \p_\varphi h_{\tau\tau}$.  Plugging
in the solutions for $h_{\mu\nu}$ from above, the final condition is
$g(\tau,r,\varphi) = 0$.  Note that the constraints imply $h \equiv \bar{g}^{\mu\nu}h_{\mu\nu} = 0$.
\section{Charge integrability}
In this appendix we show that to quadratic order around the NHEK background, the charges (\ref{gfz}) do not depend on the path of integration over metrics, $\gamma$.  Since $E_R = 0$, only $Q_{\zeta_\epsilon}[g]$ needs to be checked.  The integrability condition is
\be
\int_{\p\Sigma}\left(k_{\zeta_\epsilon}[h, g + \tilde{h}] - k_{\zeta_\epsilon}[\tilde{h}, g + h] -  k_{\zeta_\epsilon}[h-\tilde{h}, g]\right)= 0
\ee
keeping terms up to order $h\tilde{h}$.  The integrand is
\bea
&& - {1\over8} \epsilon_{\alpha\beta\mu\nu}\big[ \tilde{h}\left(
\zeta^\nu D^\mu h - \zeta^\nu D_\sigma h^{\mu\sigma}
+\half h^{\sigma\nu}(D^\mu\zeta_\sigma + D_\sigma\zeta^\mu)\right)+\zeta^\nu h^{\lambda\mu} D_\lambda \tilde{h}\\
\non   && - \zeta^\nu
(2D_\sigma \tilde{h}^\mu_{\lambda}-D^\mu
\tilde{h}_{\lambda\sigma})h^{\lambda\sigma} + \zeta^\sigma
h^{\lambda\nu}D_\sigma
\tilde{h}^\mu_{\lambda}-(h_{\sigma\lambda}\tilde{h}^{\nu\lambda}D^\mu\zeta^\sigma+h_\sigma^\nu
\tilde{h}^{\mu\lambda}D_\lambda\zeta^\sigma)\notag\\
&& - (h \leftrightarrow \tilde{h})
\big]dx^\alpha \wedge dx^\beta\notag
\eea
Using the boundary conditions (\ref{strictbc}) and the constraints $h = \tilde{h} = 0$ derived in Appendix A, the component tangent to $\p\Sigma$ vanishes on the NHEK background $\bar g_{\mu\nu}$ for $\zeta = \zeta_\e$.

\end{document}